# Describing urban evolution with the fractal parameters based on area-perimeter allometry

Yanguang Chen[1], Jiejing Wang[2*]

(1. Department of Geography, College of Urban and Environmental Sciences, Peking University, Beijing 100871, PRC. E-mail: chenyg@pku.edu.cn; 2. Department of Urban Planning and Design, The University of Hong Kong, Pokfulam Road, Hong Kong SAR. E-mail: jackiewang1120@gmail.com)

**Abstract**: The area-perimeter allometric scaling is a basic and important approach for researching fractal cities and has been studied for a long time. However, the boundary dimension of a city is always numerically overestimated by the traditional formula. An adjusting formula has been derived to revise the overestimated boundary dimension and estimate the form dimension, but the association between the global and local fractal parameters is not clear. This paper is devoted to describing the urban evolution by using the improved fractal parameters based on the area-perimeter measure relation. A system of 68 cities and towns in Yangtze River Delta, China, is taken as an example to make a case study. A discovery is that the average values of the local fractal parameters are approximately equal to the corresponding global fractal parameters of cities. This suggests that the local parameters are the decomposition of the global parameters. The novelty of this empirical study is as follows: first, the form dimension and boundary dimension are integrated to characterize the urban structure and texture; second, the global and local parameters are combined to characterize an urban system and individual cities. By illustrating how to carry out the area-perimeter scaling analysis in the case of remote sensing images with low resolution, this work suggests a possible new approach to researching fractal systems of cities.

# 1 Introduction

A scientific study should proceed first by describing how a system works and later by understanding why (Gordon, 2005; Henry, 2002). Fractal geometry is a powerful tool to describe urban systems because fractal dimension is a kind of characteristic parameters to describe scale-free phenomena (Batty, 2008; Batty and Longley, 1994; De Keersmaecker *et al*, 2003; Frankhauser, 1994; Frankhauser, 1998; Thomas *et al*, 2007; Thomas *et al*, 2010). By fractal dimension description, we can get insight into the spatial dynamics of urban evolution (Batty and Xie, 1999; Benguigui *et al*, 2001; Benguigui *et al*, 2006; White, 1998). There are two simple and convenient ways of understanding city fractals. One is measurement (see e.g. Benguigui *et al*, 2000; Feng and Chen, 2010; Chen and Wang, 2013; Shen, 2002; Thomas *et al*, 2008; White and Engelen, 1994), and the other is the geometric measure relation (see e.g. Benguigui *et al*, 2006; Chen, 2011; Longley *et al*, 1991). If a measurement result such as length, area, and density depends on the scale (the length of yardstick) to measure, we may face a fractal object. On the other hand, if the proportional relation between two measures such as urban area and urban perimeter is involved with a scaling exponent indicating a fractional parameter, we will meet a fractal phenomenon. The former way can be used to realize any types of fractals, while the latter way can be employed to identify the fractal boundary of a region. Sometimes an urbanized area is treated as a Euclidean space, but the urban boundary can be regarded as a fractal line. In this case, the boundary dimension of a city provides a good way of understanding city size and shape (Batty and Longley, 1994). So far, we have more than five empirical approaches to estimating the fractal dimension of geographical boundaries (Batty and Longley, 1988; Batty and Longley, 1989; Longley and Batty, 1989; Song *et al*, 2012; Wang *et al*, 2005).

An urban area bears an analogy to a random Koch island, thus the fractal dimension of its boundary can be estimated with the area-perimeter scaling (Batty and Longley, 1994; Feder, 1988; Mandelbrot, 1983). The geometrical measure relation between urban area and perimeter is in essence an allometric scaling relation, which is similar to the relation between urban area and population (Chen, 2010a). If the population size of a city is compared to the weight of an animal, then the urban area can be compared to the volume of the animal, and the urban perimeter, to the surface area (the area of the whole skin) (Chen, 2011). In many cases, it is difficult to investigate city population, but it is easy to measure urban area and the corresponding perimeter by using the remote-sensing images and the technology of geographical information system (GIS). The area-perimeter allometric scaling is a simple approach to revealing the spatio-temporal evolution and dynamics of urban systems. However, two problems remain to be solved. First, in many cases, it is impossible to compute the fractal dimension of each city in an urban system because of inadequate remote sensing data or the lower resolution of remote sensing images. Second, the traditional formula of the boundary dimension based on the area-perimeter scaling is not exact enough to guarantee the effective results.

The processes and patterns of urban evolution follow scaling law (Batty *et al*, 2008; Bettencourt, 2013; Bettencourt *et al*, 2007; Lobo *et al*, 2013). Fractal geometry is an effective tool to explore scaling in cities. Recent years, the new formulae for revising the boundary dimension calculations through the area-perimeter scaling have been proposed (Chen, 2013). By means of these formulae, we can correct the errors of the boundary dimension and estimate the form dimension of cities. Using the boundary dimension and form dimension, we can characterize the spatio-temporal evolution of urban systems. This paper is devoted to researching urban evolution properties and trends using the fractal dimension sets based on the area-perimeter allometric scaling. It tries to solve several problems such as: how to combine the global fractal parameters with the local fractal parameters for spatial analysis; how to make use of remote sensing images of low resolution for urban fractal studies; how to understand the influence of urban sprawl on fractal dimension change. The rest of the work is organized as follows. In Section 2, several new formulae of fractal dimension estimation are introduced and clarified for spatial-temporal analysis of urban form and growth. In Section 3, the new models and formulae are applied to the system of cities and towns in Yangtze River Delta, China, to make an empirical analysis. In Section 4, the related questions are discussed. Based on the discussion, the article reaches its conclusions.

## 2 Fractal parameters

### 2.1 Basic formulae

In theory, a city figure can be divided into two parts: one is the urban boundary, and the other is the urban area within the boundary. The former is termed *urban envelope* (E) and can be described with the boundary dimension (Longley *et al*, 1991), while the latter is named *urban area* (A) and can be characterized with form dimension (Batty and Longley, 1994; Chen, 2010b). The form dimension is a *structural dimension*, while the boundary dimension is a *textural dimension* (Addison, 1997; Chen, 2011; Kaye, 1989). In technique, the urban area can be regarded as a Euclidean plane with a dimension $d=2$, and accordingly, the urban boundary is treated as a fractal line (Batty and Longley, 1994; Chen, 2013). Thus, the boundary dimension represents fractal dimension of urban shape, and can be estimated with the regression analysis based on the method of the ordinary least squares (OLS). By the fractal measure relation (Feder, 1988; Mandelbrot, 1983; Takayasu, 1990), the urban area and perimeter follow a power law as below

$$(\frac{P}{k})^{1/D_l} = A^{1/2}, \qquad (1)$$

where $P$ refers to the perimeter of the urban envelope, and $A$ to the corresponding urban area. As for the parameters, $D_l$ denotes the boundary dimension, i.e., the fractal dimension of urban boundary, and $k$ is related with the proportionality coefficient. In this context, the boundary dimension $D_l$ should be termed *initial boundary dimension*. Equation (1) is in fact an allometric scaling relation, and the scaling exponent is

$$b = \frac{D_l}{d} = \frac{D_l}{2}, \qquad (2)$$

where $d=2$ denotes the Euclidean dimension of the embedding space of urban form. The boundary dimension is often estimated by the formula: $D_l=2b$. From equation (1) it follows

$$D_l = \frac{2\ln(P/k)}{\ln(A)}, \qquad (3)$$

which is an approximate formula of the boundary dimension estimation. By analogy with squares and empirical analysis, the proportionality parameter is always taken as $k=4$ (Chang, 1996; Chang and Wu, 1998). Thus, equation (2) provides a simple approach to estimating the boundary dimension especially when spatial data are short to computing the fractal dimension.

## 2.2 Adjusting formulae

It is easy to evaluate the boundary dimension of a fractal region such as the Koch island and urban envelope. However, the method based on the geometric measure relation always overestimates the fractal dimension value (Chen, 2010a). In order to lessen the errors resulting from equation (1) and (2), a formula is derived as follows (Chen, 2013):

$$D_b = \frac{1+D_l}{2}, \quad (4)$$

where $D_b$ represents the *revised boundary dimension* of a city (a textural dimension). Equation (4) suggests a linear relation between $D_l$ and $D_b$. If we calculate the value of $D_l$ using equation (3) or log-linear regression analysis, we can estimate the $D_b$ value by means of equation (4).

A real city is a complex spatial system, and urban area do not correspond to a 2-dimensional region. In this case, equation (1) is not enough to describe the geometric measure relation between urban area and perimeter. According to the studies of Cheng (1995), Imre (2006), and Imre and Bogaert (2004), equation (1) can be generalized as follows

$$(kP)^{1/D_b} = A^{1/D_f}, \quad (5)$$

where $D_f$ denotes the fractal dimension of urban form within the urban envelope (a structural dimension). The form dimension $D_f$ can be estimated with the following formula (Chen, 2013)

$$D_f = 1 + \frac{1}{D_l}, \quad (6)$$

which suggests a hyperbolic relation between $D_l$ and $D_f$. From equation (4) and (6) it follows

$$\frac{1}{D_f} = 1 - \frac{1}{2D_b}, \quad (7)$$

which suggests a hyperbolic relation between $D_b$ and $D_f$. Accordingly, $1/D_b = 2-2/D_f$. If $D_b=1$, then we have $D_f=2$, and *vice versa*. If so, we will have a Euclidean object. Based on equation (5), the allometric scaling exponent expressed by equation (2) can be rewritten as

$$\sigma = \frac{D_b}{D_f}, \quad (8)$$

where $\sigma$ refers to the revised scaling exponent of the area-perimeter allometry. An allometric exponent is usually a ratio of one fractal dimension to another fractal dimension. In many cases, it is the allometric scaling exponents rather than fractal dimensions that play an important role in spatial analysis of urban systems (Chen, 2010a; Chen and Jiang, 2009; Luo and Chen, 2014).

Using the mathematical models, allometric scaling relations, and the fractal parameter formulae, we can describe and analyze the spatial development and evolution of urban systems in the real world. For an urban system, the fractal dimensions and the related allometric scaling exponents can be classified as global fractal parameters and local fractal parameters. The global fractal parameters can be reckoned with the cross-sectional data and used to describe a system of cities as a whole, while the local fractal parameters can be figured out by using the data of individual cities and used to describe each city as an element in the urban system.

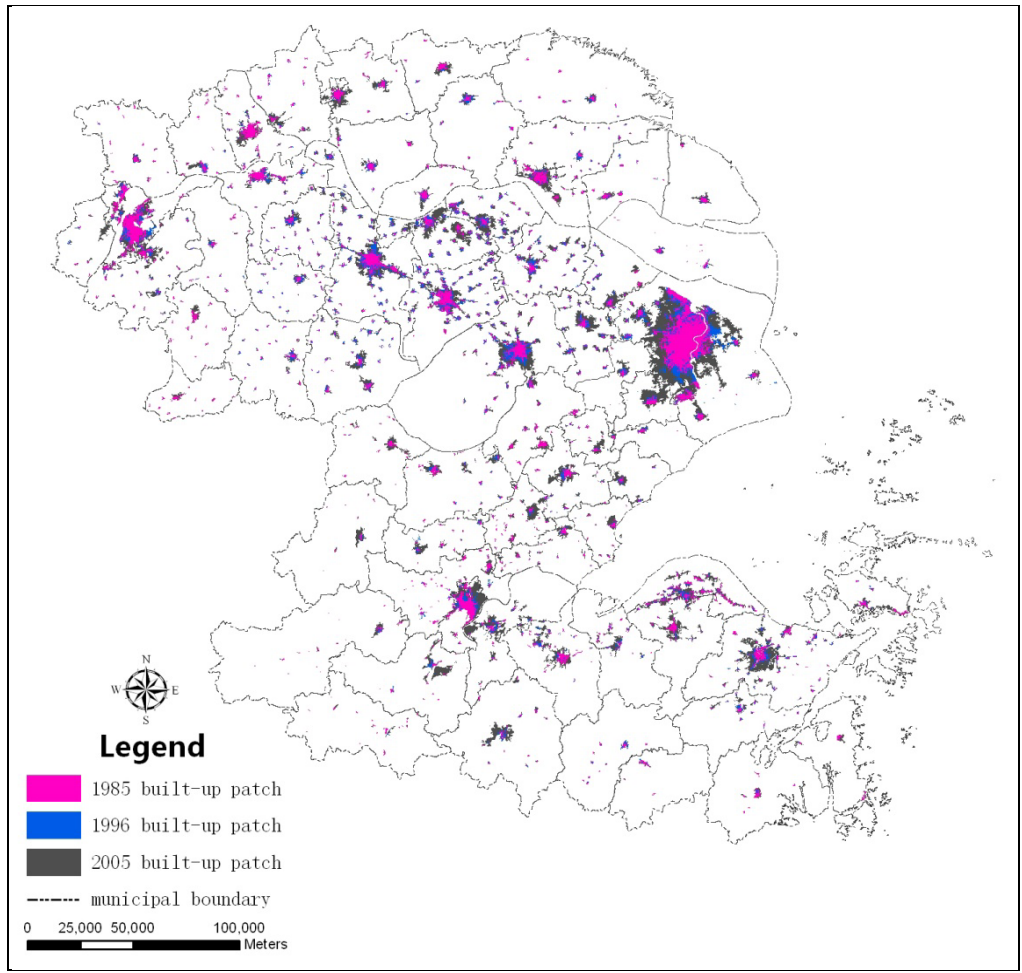

**Figure 1 A sketch map of the 68 cities and towns in Yangtze River delta, China (1985-2005)**

# 3 Empirical analysis

## 3.1 Study area and method

The area-perimeter scaling and the adjusting formulae of fractal dimension can be applied to the actual cities by means of remote sensing data. Yangtze River Delta in China is taken as a study area to make an empirical analysis. The region includes 68 cities and towns, which can be regarded as urban system (Figure 1). The remote sensing images used in this research came from National Aeronautics and Space Administration (NASA), including Landsat MSS, TM and ETM images from 1985, 1996 and 2005, which were download from the Earth Science Data Interface (ESDI) at the Global Land Cover Facility (GLCF) center of University of Maryland (http://glcfapp.umiacs.umd.edu:8080/esdi). These images were first transformed to the GCS_WGS_1984 geographic coordinate system and Asia_Lambert_Conformal_Conic projected coordinate system. Then the supervised classification method (SCM) was employed to extract the built-up area and the boundary of each city in a given year with ERDAS IMAGINE software. And then, we modified the result manually in ESRI ArcGIS to ensure better accuracy. After extracting the built-up areas, it is convenient to obtain the urban boundaries in ArcGIS. Using the *Intersect* function of ArcToolbox, we could extract the built-up area which falls into each urban envelope, and then calculate areal and perimetric values through the *Calculate Geometry* function in the Attribute Table. Finally, we have 3 datasets of urban areas and perimeters for the 68 cities and towns in three years (Table 1).

If the observational data have been processed, the procedure of parameter estimation is as follows:

**Step 1: preliminary estimation of fractal parameters**. Using the geometric measure relation, scatterplots, and the least squares calculations, we can evaluate the initial boundary dimension $D_l$ and the proportionality coefficient $k$. the model and formulae include equations (1) and (2).

**Step 2: revision of the global fractal parameters**. Using equations (4) and (6), we can transform the initial boundary dimension $D_l$ into the revised boundary dimension $D_b$ and form dimension $D_f$, which are used to describe the urban system in the study area.

**Step 3: estimation of the local fractal parameters**. Using the approximate formula, equation (3), and the $k$ values from step 1, we can estimate the initial boundary dimension $D_l$ for each city.

**Step 4: revision of the local fractal parameters**. Using equations (4) and (6), we can transform the initial local boundary dimension $D_l$ into the revised boundary dimension $D_b$ and form dimension $D_f$, which are used to describe the individual cities in the study area.

**Step 5: related calculations**. Fractal parameters can be associated with traditional spatial measurements. For example, the local boundary dimension is a function of the compactness of urban form; therefore, the compactness ratio can be figured out with urban area and perimeter.

Table 1 The urban area and perimeter datasets of the cities and towns in Yangtze River Delta, China (1985-2005)

| City/Town | 1985 | | 1996 | | 2005 | |
| --- | --- | --- | --- | --- | --- | --- |
| | Area $A$ | Perimeter $P$ | Area $A$ | Perimeter $P$ | Area $A$ | Perimeter $P$ |
| Shanghai | 492.52 | 383.30 | 657.73 | 629.58 | 1368.11 | 1095.44 |
| Qingpu | 5.23 | 33.56 | 9.36 | 29.78 | 25.93 | 59.29 |
| Chongming | 4.89 | 24.51 | 6.29 | 22.85 | 9.69 | 27.51 |
| Songjiang | 12.88 | 49.74 | 15.46 | 26.74 | 71.93 | 83.81 |
| Jinshan | 4.43 | 20.53 | 5.72 | 21.83 | 27.36 | 35.15 |
| Nanjing | 113.24 | 286.55 | 171.48 | 257.13 | 311.11 | 418.07 |
| Jiangning | 7.72 | 28.23 | 17.62 | 57.84 | 41.85 | 91.62 |
| Jiangpu | 3.70 | 26.97 | 8.29 | 49.04 | 15.84 | 50.94 |
| Liuhe | 4.67 | 22.44 | 7.95 | 26.16 | 16.17 | 34.78 |
| Lishui | 9.46 | 43.18 | 9.46 | 37.82 | 21.04 | 53.97 |
| Gaochun | 4.47 | 19.26 | 4.47 | 19.26 | 11.32 | 37.98 |
| Wuxi | 60.54 | 152.58 | 98.23 | 137.39 | 149.01 | 192.22 |
| Jiangyin | 15.51 | 46.77 | 25.75 | 69.01 | 95.17 | 199.15 |
| Yixing | 5.13 | 29.33 | 11.37 | 37.58 | 38.09 | 78.15 |
| Changzhou | 64.98 | 156.09 | 114.00 | 207.79 | 205.62 | 275.72 |
| Liyang | 3.79 | 22.46 | 12.71 | 56.49 | 18.95 | 64.27 |
| Jintan | 4.69 | 25.72 | 10.91 | 41.98 | 19.69 | 46.87 |
| Suzhou | 54.19 | 137.89 | 104.30 | 166.28 | 197.10 | 180.47 |
| Changshu | 11.92 | 67.02 | 25.26 | 66.09 | 58.60 | 127.81 |
| Zhangjiagang | 6.70 | 37.78 | 18.42 | 47.52 | 70.19 | 112.34 |
| Kunshan | 8.85 | 45.19 | 18.92 | 39.67 | 62.45 | 79.94 |
| Wujiang | 6.47 | 28.46 | 7.92 | 21.45 | 18.73 | 34.46 |
| Taicang | 7.57 | 35.55 | 11.29 | 29.53 | 38.82 | 55.76 |
| Nantong | 27.94 | 53.85 | 46.99 | 109.35 | 114.53 | 226.23 |
| Hai'an | 10.70 | 36.20 | 12.99 | 32.86 | 29.97 | 57.21 |
| Rudong | 4.09 | 16.44 | 7.06 | 19.32 | 14.23 | 31.38 |
| Haimen | 5.81 | 19.05 | 7.55 | 25.01 | 15.53 | 44.87 |
| Qidong | 6.35 | 19.33 | 8.57 | 23.42 | 20.26 | 51.45 |
| Rugao | 5.74 | 29.75 | 15.42 | 60.47 | 23.25 | 53.01 |
| Yangzhou | 38.53 | 112.91 | 44.59 | 154.97 | 75.57 | 154.09 |
| Jiangdu | 7.60 | 29.05 | 10.36 | 41.29 | 31.05 | 85.12 |
| Yizheng | 8.73 | 47.51 | 13.86 | 38.58 | 24.37 | 49.05 |
| Taizhou | 17.80 | 58.73 | 21.51 | 64.12 | 70.61 | 169.03 |
| Jiangyan (Tai) | 6.48 | 28.73 | 8.72 | 38.09 | 24.98 | 80.58 |
| Taixing | 9.45 | 38.95 | 15.11 | 40.68 | 23.77 | 65.96 |
| Jingjiang | 11.07 | 49.63 | 16.11 | 31.82 | 28.23 | 78.59 |
| Zhenjiang | 29.67 | 109.18 | 44.83 | 130.62 | 67.51 | 186.67 |
| Jurong | 3.54 | 25.73 | 6.74 | 27.58 | 10.06 | 33.42 |
| Yangzhong | 3.65 | 19.13 | 6.28 | 21.64 | 12.25 | 45.72 |
| Danyang | 7.33 | 38.86 | 15.09 | 42.77 | 33.02 | 67.58 |
| Hangzhou | 67.49 | 160.88 | 103.41 | 213.64 | 200.66 | 184.77 |
| Tonglu | 1.98 | 12.56 | 2.78 | 14.11 | 7.43 | 30.32 |
| Fuyang | 2.94 | 15.79 | 7.82 | 34.39 | 55.75 | 83.71 |
| Lin'an | 3.86 | 25.85 | 6.26 | 26.81 | 16.25 | 34.92 |
| Yuhang | 1.47 | 10.33 | 6.04 | 24.35 | 17.88 | 40.72 |
| Xiaoshan | 4.99 | 21.17 | 16.76 | 69.84 | 96.49 | 178.18 |
| Ningbo | 29.93 | 82.98 | 50.54 | 99.57 | 193.81 | 219.05 |
| Xiangshan | 3.62 | 21.26 | 2.54 | 11.42 | 7.39 | 20.01 |
| Ninghai | 5.61 | 20.31 | 7.26 | 22.15 | 12.51 | 23.07 |
| Yuyao | 14.40 | 64.56 | 15.69 | 87.89 | 63.83 | 157.04 |
| Cixi | 4.67 | 24.09 | 17.69 | 54.75 | 67.35 | 146.38 |
| Fenghua | 3.29 | 18.32 | 8.91 | 34.51 | 18.95 | 82.31 |
| Jiaxing | 11.71 | 46.17 | 19.47 | 45.62 | 67.29 | 98.29 |
| Jiashan | 4.31 | 22.66 | 4.72 | 13.57 | 23.27 | 45.71 |
| Haiyan | 3.97 | 22.85 | 6.09 | 20.09 | 21.49 | 29.12 |
| Haining | 5.64 | 22.75 | 8.27 | 24.67 | 23.34 | 57.09 |
| Pinghu | 3.04 | 15.19 | 6.26 | 21.08 | 24.78 | 40.35 |
| Tongxiang | 3.29 | 18.35 | 5.55 | 20.27 | 34.34 | 43.68 |
| Huzhou | 10.43 | 28.61 | 14.19 | 31.84 | 30.15 | 50.59 |
| Deqing | 1.49 | 5.77 | 3.05 | 9.87 | 22.52 | 33.88 |
| Changxing | 6.81 | 30.88 | 7.34 | 30.28 | 23.24 | 43.68 |
| Anji | 2.14 | 13.21 | 3.53 | 18.39 | 23.14 | 50.29 |
| Shaoxing | 20.21 | 49.00 | 23.08 | 64.55 | 57.23 | 142.19 |
| Chengxian | 3.38 | 16.11 | 7.59 | 22.36 | 8.08 | 28.06 |
| Xinchang | 5.87 | 24.32 | 6.75 | 23.01 | 7.86 | 28.05 |
| Zhuji | 3.57 | 24.14 | 12.82 | 55.66 | 65.86 | 110.29 |
| Shangyu | 1.68 | 13.67 | 7.95 | 42.16 | 28.77 | 72.33 |
| Zhoushan | 7.28 | 46.57 | 8.68 | 42.44 | 14.76 | 47.78 |

## 3.2 Results

The regression analysis based on OLS can be employed to estimate the allometric parameters and the global fractal parameters. The log-log scatterplots show that the relations between urban area and perimeter follow the allometric scaling law (Figure 2). The slopes of the trendlines give the values of the scaling exponent $b$ in different years. Taking natural logarithms on both sides of equation (1) yields a linear relation as below

$$\ln P = \ln k + \frac{D_l}{2} \ln A = C + b \ln A, \qquad (9)$$

where $C=\ln k$, $b=D_l/2$. By means of the least squares calculations, we can fit equation (9) to the datasets displayed in Table 1. The results including the values of the proportionality coefficient $k$, scaling exponent $b$, and the goodness of fit $R^2$ are tabulated as below (Table 2). Using equations (2), (4), and (6), we can further estimate the boundary dimension $D_l$, the revised boundary dimension $D_b$, and the form dimension $D_f$. The values of boundary dimension $D_l$ range from 1.4 and 1.5. Accordingly, the revised boundary dimension $D_b$ is about 1.2, and the corresponding form dimension $D_f$ is around 1.7.

Table 2 The global fractal parameters of the cities and towns in Yangtze River Delta, China (1985-2005)

| Year | Original results | | | | Revised results | |
|---|---|---|---|---|---|---|
| | $k$ | $b$ | $D_l=2b$ | $R^2$ | $D_b=(1+D_l)/2$ | $D_f=1+1/D_l$ |
| 1985 | 8.0340 | 0.6963 | 1.3926 | 0.9282 | 1.1963 | 1.7181 |
| 1996 | 6.1128 | 0.7387 | 1.4774 | 0.9144 | 1.2387 | 1.6769 |
| 2005 | 5.4610 | 0.7266 | 1.4532 | 0.8879 | 1.2266 | 1.6882 |
| Average | 6.5359 | 0.7205 | 1.4411 | 0.9102 | 1.2205 | 1.6944 |

If we had the remote sensing images with enough high resolution of all the cities, we would calculate the form and boundary dimension for each city (Chen and Wang, 2013; Wang et al, 2005). If so, the local fractal parameters could be evaluated with the regression analysis or other algorithms such as the major axis (MA) method. Unfortunately, we have only the images of the region instead of each city. In this case, the local parameters can be estimated by equation (3). The key is to determine the value of *k*. There exist two possible approaches to estimating the *k* value. One is empirical method. As indicated above, a fixed *k*=4 is proposed by empirical analysis (Chang, 1996; Chang and Wu, 1998). However, if we take *k*=4, the boundary dimension of the cities in our study area will be significantly overestimated. For example, for 1985, the average $D_l$ value of the 68 cities and towns is about 2.258, which is greater than the Euclidean dimension of the embedding space and thus unreasonable. Generally speaking, the boundary dimension comes between 1 and 1.5, and the average value is often near 1.25 (Chen, 2010a; Chen, 2011). The other is the regression method. By the least squares calculation of cross-sectional datasets, we can estimate the *k* values using equation (9). This is an unfixed parameter: the *k* value changes over time. Using the variable *k* value to replace the fixed *k* value in equation (3), we can estimate the boundary dimension $D_l$ of each city. Then, using equations (4) and (6), we can further estimate the revised boundary dimension $D_b$ and the corresponding form dimension $D_f$ for each city. All these results represent the local fractal parameters of the urban system in the study region (Table 3).

Table 3 The local fractal parameters of the cities and towns in Yangtze River Delta, China (1985-2005)

| City | 1985 | | | | 1996 | | | | 2005 | | | |
|---|---|---|---|---|---|---|---|---|---|---|---|---|
| | $D_l$ | $D_b$ | $D_f$ | Co | $D_l$ | $D_b$ | $D_f$ | Co | $D_l$ | $D_b$ | $D_f$ | Co |
| Shanghai | 1.247 | 1.123 | 1.802 | 0.205 | 1.429 | 1.214 | 1.700 | 0.144 | 1.468 | 1.234 | 1.681 | 0.120 |
| Qingpu | 1.728 | 1.364 | 1.579 | 0.242 | 1.416 | 1.208 | 1.706 | 0.364 | 1.465 | 1.233 | 1.683 | 0.304 |
| Chongming | 1.405 | 1.203 | 1.711 | 0.320 | 1.434 | 1.217 | 1.697 | 0.389 | 1.424 | 1.212 | 1.702 | 0.401 |
| Songjiang | 1.427 | 1.213 | 1.701 | 0.256 | 1.078 | 1.039 | 1.928 | 0.521 | 1.277 | 1.139 | 1.783 | 0.359 |
| Jinshan | 1.261 | 1.130 | 1.793 | 0.363 | 1.460 | 1.230 | 1.685 | 0.388 | 1.125 | 1.063 | 1.889 | 0.528 |
| Nanjing | 1.511 | 1.256 | 1.662 | 0.132 | 1.454 | 1.227 | 1.688 | 0.181 | 1.511 | 1.256 | 1.662 | 0.150 |
| Jiangning | 1.230 | 1.115 | 1.813 | 0.349 | 1.567 | 1.283 | 1.638 | 0.257 | 1.510 | 1.255 | 1.662 | 0.250 |
| Jiangpu | 1.851 | 1.426 | 1.540 | 0.253 | 1.969 | 1.484 | 1.508 | 0.208 | 1.617 | 1.308 | 1.619 | 0.277 |
| Liuhe | 1.333 | 1.166 | 1.750 | 0.341 | 1.403 | 1.201 | 1.713 | 0.382 | 1.330 | 1.165 | 1.752 | 0.410 |
| Lishui | 1.497 | 1.248 | 1.668 | 0.253 | 1.622 | 1.311 | 1.616 | 0.288 | 1.504 | 1.252 | 1.665 | 0.301 |
| Gaochun | 1.168 | 1.084 | 1.856 | 0.389 | 1.533 | 1.266 | 1.652 | 0.389 | 1.598 | 1.299 | 1.626 | 0.314 |
| Wuxi | 1.435 | 1.217 | 1.697 | 0.181 | 1.357 | 1.178 | 1.737 | 0.256 | 1.423 | 1.212 | 1.703 | 0.225 |
| Jiangyin | 1.285 | 1.143 | 1.778 | 0.298 | 1.492 | 1.246 | 1.670 | 0.261 | 1.579 | 1.289 | 1.633 | 0.174 |
| Yixing | 1.584 | 1.292 | 1.631 | 0.274 | 1.494 | 1.247 | 1.669 | 0.318 | 1.462 | 1.231 | 1.684 | 0.280 |
| Changzhou | 1.422 | 1.211 | 1.703 | 0.183 | 1.489 | 1.245 | 1.672 | 0.182 | 1.473 | 1.236 | 1.679 | 0.184 |
| Liyang | 1.543 | 1.272 | 1.648 | 0.307 | 1.749 | 1.375 | 1.572 | 0.224 | 1.676 | 1.338 | 1.597 | 0.240 |
| Jintan | 1.506 | 1.253 | 1.664 | 0.298 | 1.613 | 1.306 | 1.620 | 0.279 | 1.443 | 1.221 | 1.693 | 0.336 |
| Suzhou | 1.424 | 1.212 | 1.702 | 0.189 | 1.422 | 1.211 | 1.703 | 0.218 | 1.324 | 1.162 | 1.755 | 0.276 |
| Changshu | 1.712 | 1.356 | 1.584 | 0.183 | 1.474 | 1.237 | 1.678 | 0.270 | 1.549 | 1.275 | 1.646 | 0.212 |
| Zhangjiagang | 1.628 | 1.314 | 1.614 | 0.243 | 1.408 | 1.204 | 1.710 | 0.320 | 1.423 | 1.211 | 1.703 | 0.264 |
| Kunshan | 1.584 | 1.292 | 1.631 | 0.233 | 1.272 | 1.136 | 1.786 | 0.389 | 1.298 | 1.149 | 1.770 | 0.350 |
| Wujiang | 1.355 | 1.177 | 1.738 | 0.317 | 1.213 | 1.107 | 1.824 | 0.465 | 1.257 | 1.129 | 1.795 | 0.445 |
| Taicang | 1.469 | 1.235 | 1.681 | 0.274 | 1.300 | 1.150 | 1.769 | 0.403 | 1.270 | 1.135 | 1.787 | 0.396 |
| Nantong | 1.143 | 1.071 | 1.875 | 0.348 | 1.498 | 1.249 | 1.667 | 0.222 | 1.571 | 1.285 | 1.637 | 0.168 |
| Hai'an | 1.270 | 1.135 | 1.787 | 0.320 | 1.312 | 1.156 | 1.762 | 0.389 | 1.382 | 1.191 | 1.724 | 0.339 |

| City | | | | | | | | | | | | |
|---|---|---|---|---|---|---|---|---|---|---|---|---|
| Rudong | 1.017 | 1.008 | 1.984 | 0.436 | 1.178 | 1.089 | 1.849 | 0.488 | 1.317 | 1.158 | 1.759 | 0.426 |
| Haimen | 0.981 | 0.991 | 2.019 | 0.449 | 1.394 | 1.197 | 1.717 | 0.389 | 1.536 | 1.268 | 1.651 | 0.311 |
| Qidong | 0.950 | 0.975 | 2.053 | 0.462 | 1.251 | 1.125 | 1.800 | 0.443 | 1.491 | 1.246 | 1.671 | 0.310 |
| Rugao | 1.498 | 1.249 | 1.667 | 0.285 | 1.675 | 1.338 | 1.597 | 0.230 | 1.445 | 1.222 | 1.692 | 0.322 |
| Yangzhou | 1.448 | 1.224 | 1.691 | 0.195 | 1.703 | 1.351 | 1.587 | 0.153 | 1.544 | 1.272 | 1.647 | 0.200 |
| Jiangdu | 1.267 | 1.134 | 1.789 | 0.336 | 1.634 | 1.317 | 1.612 | 0.276 | 1.599 | 1.299 | 1.625 | 0.232 |
| Yizheng | 1.640 | 1.320 | 1.610 | 0.220 | 1.402 | 1.201 | 1.713 | 0.342 | 1.375 | 1.187 | 1.727 | 0.357 |
| Taizhou | 1.382 | 1.191 | 1.724 | 0.255 | 1.532 | 1.266 | 1.653 | 0.256 | 1.613 | 1.306 | 1.620 | 0.176 |
| Jiangyan (Tai) | 1.364 | 1.182 | 1.733 | 0.314 | 1.690 | 1.345 | 1.592 | 0.275 | 1.673 | 1.336 | 1.598 | 0.220 |
| Taixing | 1.406 | 1.203 | 1.711 | 0.280 | 1.396 | 1.198 | 1.716 | 0.339 | 1.573 | 1.286 | 1.636 | 0.262 |
| Jingjiang | 1.515 | 1.257 | 1.660 | 0.238 | 1.187 | 1.094 | 1.842 | 0.447 | 1.597 | 1.298 | 1.626 | 0.240 |
| Zhenjiang | 1.539 | 1.270 | 1.650 | 0.177 | 1.610 | 1.305 | 1.621 | 0.182 | 1.677 | 1.338 | 1.596 | 0.156 |
| Jurong | 1.842 | 1.421 | 1.543 | 0.259 | 1.579 | 1.290 | 1.633 | 0.334 | 1.569 | 1.285 | 1.637 | 0.336 |
| Yangzhong | 1.340 | 1.170 | 1.746 | 0.354 | 1.376 | 1.188 | 1.727 | 0.411 | 1.696 | 1.348 | 1.590 | 0.271 |
| Danyang | 1.583 | 1.291 | 1.632 | 0.247 | 1.434 | 1.217 | 1.698 | 0.322 | 1.439 | 1.219 | 1.695 | 0.301 |
| Hangzhou | 1.423 | 1.212 | 1.703 | 0.181 | 1.532 | 1.266 | 1.653 | 0.169 | 1.328 | 1.164 | 1.753 | 0.272 |
| Tonglu | 1.308 | 1.154 | 1.764 | 0.397 | 1.636 | 1.318 | 1.611 | 0.419 | 1.709 | 1.355 | 1.585 | 0.319 |
| Fuyang | 1.253 | 1.127 | 1.798 | 0.385 | 1.680 | 1.340 | 1.595 | 0.288 | 1.358 | 1.179 | 1.736 | 0.316 |
| Lin'an | 1.730 | 1.365 | 1.578 | 0.269 | 1.612 | 1.306 | 1.620 | 0.331 | 1.331 | 1.165 | 1.751 | 0.409 |
| Yuhang | 1.305 | 1.152 | 1.766 | 0.416 | 1.537 | 1.269 | 1.651 | 0.358 | 1.393 | 1.197 | 1.718 | 0.368 |
| Xiaoshan | 1.206 | 1.103 | 1.830 | 0.374 | 1.728 | 1.364 | 1.579 | 0.208 | 1.525 | 1.263 | 1.656 | 0.195 |
| Ningbo | 1.374 | 1.187 | 1.728 | 0.234 | 1.423 | 1.211 | 1.703 | 0.253 | 1.402 | 1.201 | 1.713 | 0.225 |
| Xiangshan | 1.513 | 1.256 | 1.661 | 0.317 | 1.341 | 1.170 | 1.746 | 0.495 | 1.299 | 1.149 | 1.770 | 0.482 |
| Ninghai | 1.076 | 1.038 | 1.930 | 0.413 | 1.299 | 1.149 | 1.770 | 0.431 | 1.141 | 1.070 | 1.877 | 0.543 |
| Yuyao | 1.563 | 1.281 | 1.640 | 0.208 | 1.937 | 1.468 | 1.516 | 0.160 | 1.616 | 1.308 | 1.619 | 0.180 |
| Cixi | 1.425 | 1.213 | 1.702 | 0.318 | 1.526 | 1.263 | 1.655 | 0.272 | 1.562 | 1.281 | 1.640 | 0.199 |
| Fenghua | 1.384 | 1.192 | 1.722 | 0.351 | 1.583 | 1.291 | 1.632 | 0.307 | 1.844 | 1.422 | 1.542 | 0.187 |
| Jiaxing | 1.421 | 1.211 | 1.704 | 0.263 | 1.354 | 1.177 | 1.739 | 0.343 | 1.373 | 1.187 | 1.728 | 0.296 |
| Jiashan | 1.420 | 1.210 | 1.704 | 0.325 | 1.028 | 1.014 | 1.973 | 0.568 | 1.350 | 1.175 | 1.741 | 0.374 |
| Haiyan | 1.516 | 1.258 | 1.660 | 0.309 | 1.317 | 1.159 | 1.759 | 0.435 | 1.091 | 1.046 | 1.916 | 0.564 |
| Haining | 1.203 | 1.102 | 1.831 | 0.370 | 1.321 | 1.160 | 1.757 | 0.413 | 1.490 | 1.245 | 1.671 | 0.300 |
| Pinghu | 1.146 | 1.073 | 1.873 | 0.407 | 1.350 | 1.175 | 1.741 | 0.421 | 1.246 | 1.123 | 1.803 | 0.437 |
| Tongxiang | 1.387 | 1.194 | 1.721 | 0.350 | 1.399 | 1.199 | 1.715 | 0.412 | 1.176 | 1.088 | 1.850 | 0.476 |
| Huzhou | 1.083 | 1.042 | 1.923 | 0.400 | 1.244 | 1.122 | 1.804 | 0.419 | 1.307 | 1.154 | 1.765 | 0.385 |
| Deqing* | -1.660 | -0.330 | 0.398 | 0.750 | 0.859 | 0.930 | 2.164 | 0.627 | 1.172 | 1.086 | 1.853 | 0.497 |
| Changxing | 1.404 | 1.202 | 1.712 | 0.300 | 1.605 | 1.303 | 1.623 | 0.317 | 1.322 | 1.161 | 1.756 | 0.391 |
| Anji | 1.307 | 1.154 | 1.765 | 0.393 | 1.747 | 1.373 | 1.573 | 0.362 | 1.413 | 1.207 | 1.708 | 0.339 |
| Shaoxing | 1.203 | 1.101 | 1.831 | 0.325 | 1.502 | 1.251 | 1.666 | 0.264 | 1.611 | 1.305 | 1.621 | 0.189 |
| Chengxian | 1.143 | 1.071 | 1.875 | 0.405 | 1.280 | 1.140 | 1.781 | 0.437 | 1.567 | 1.283 | 1.638 | 0.359 |
| Xinchang | 1.252 | 1.126 | 1.799 | 0.353 | 1.388 | 1.194 | 1.720 | 0.400 | 1.587 | 1.294 | 1.630 | 0.354 |
| Zhuji | 1.729 | 1.365 | 1.578 | 0.277 | 1.732 | 1.366 | 1.577 | 0.228 | 1.435 | 1.218 | 1.697 | 0.261 |
| Shangyu | 2.049 | 1.525 | 1.488 | 0.336 | 1.863 | 1.431 | 1.537 | 0.237 | 1.538 | 1.269 | 1.650 | 0.263 |
| Zhoushan | 1.770 | 1.385 | 1.565 | 0.205 | 1.793 | 1.397 | 1.558 | 0.246 | 1.611 | 1.306 | 1.621 | 0.285 |
| **Average** | **1.364** | **1.182** | **1.707** | **0.307** | **1.472** | **1.236** | **1.693** | **0.330** | **1.454** | **1.227** | **1.696** | **0.307** |

**Note**: The results of Deqing in 1985 are outliers. The symbol "*Co*" denotes compactness ratio of urban form.

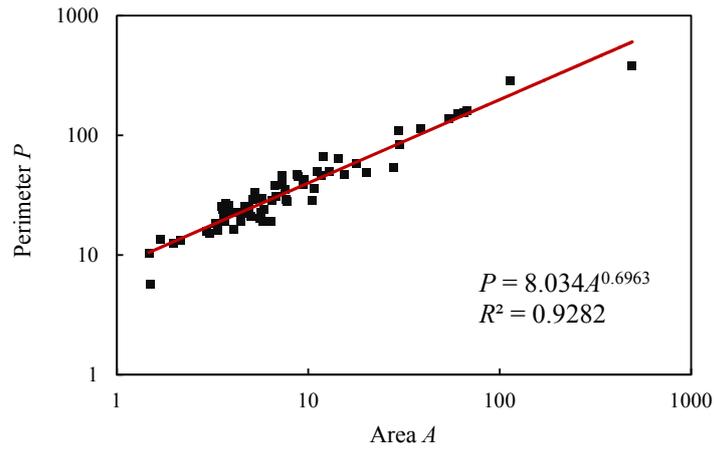

a. 1985

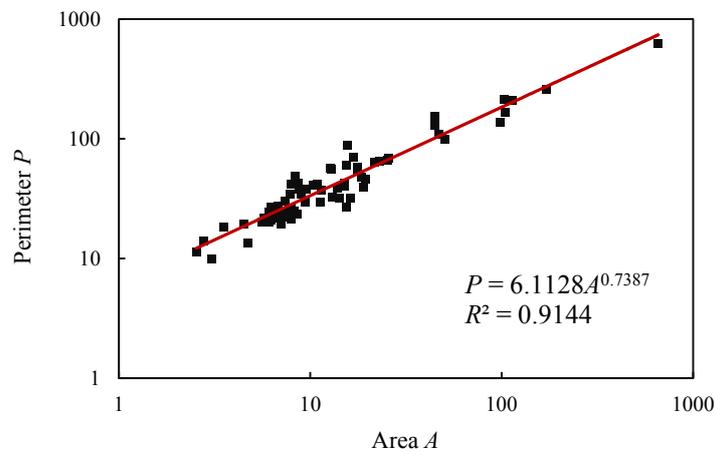

b. 1996

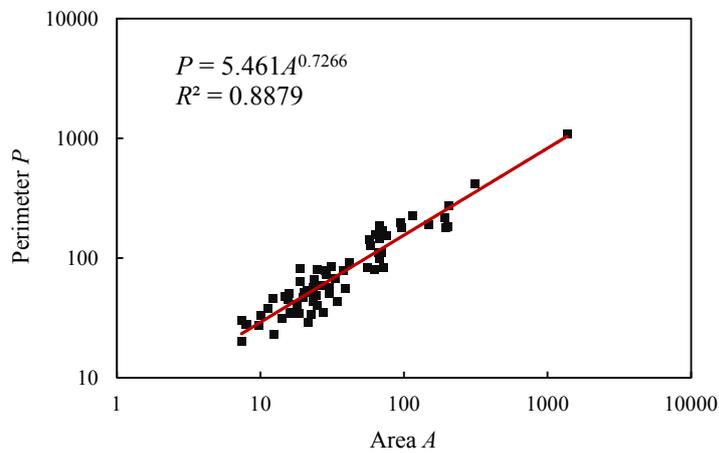

c. 2005

**Figure 2 The allometric scaling relations between urban area and perimeter of the cities and towns in Yangtze River Delta, China (1985-2005)**

A finding is that the average values of the local fractal parameters are approximately equal to the corresponding global fractal parameters of the whole cities and towns. Comparing the means of the fractal dimensions listed in Table 3 with the related parameter values displayed in Table 2, we can find the consistency of the global estimation with the local average (Table 4). The global parameter values can be used to analyze the development of the system of cities and towns at the macro level (the whole), while the local parameter values can be employed to analyze the evolution of the individual cities at the micro level (the parts or elements).

Table 4 A comparison between the global fractal parameters and the average local fractal parameters of the cities and towns in Yangtze River Delta, China (1985-2005)

| Year | Global estimation | | | Local average | | |
|---|---|---|---|---|---|---|
| | $D_l$ | $D_b$ | $D_f$ | $D_l$ | $D_b$ | $D_f$ |
| **1985** | 1.3926 | 1.1963 | 1.7181 | 1.3636 | 1.1818 | 1.7069 |
| **1996** | 1.4774 | 1.2387 | 1.6769 | 1.4722 | 1.2361 | 1.6933 |
| **2005** | 1.4532 | 1.2266 | 1.6882 | 1.4545 | 1.2272 | 1.6957 |
| **Average** | 1.4411 | 1.2205 | 1.6944 | 1.4301 | 1.2150 | 1.6986 |

### 3.3 Global and local analysis

Generally speaking, a mathematical model reflects the macro structural properties of a system, while the model parameters reflect the micro interaction of elements. It is necessary to examine the mathematical expressions of area-perimeter relations and the corresponding fractal parameters. Analyzing the model forms and parameter values results in a number of new findings.

First, the allometric scaling degenerated from power law to linear relations during from 1985 to 2005. A real allometric scaling relation takes on a double logarithmic relation, but it sometimes degenerates and changes to single logarithmic relations including exponential relation and logarithmic relation (semi-degeneration), or even to a linear relation (full degeneration) (Chen, 1995). In 1985, the area-perimeter relation followed the power law, and the goodness of fit ($R^2$) of the power function was significantly higher than those of the linear function and the single logarithmic functions. However, in 1996, the case was different, and the goodness of fit of the linear function is a little higher than that of the power function. In 2005, the goodness of fit of the linear function is significantly higher than that of the power function (Table 5). Fractals suggest the optimum structure of nature. The degeneration of the fractal measure relation indicates some disorder problem in the processes of urbanization and urban evolution. If the power law degenerates into a linear relation, we can treat it as a quasi-allometric scaling in light of fractal theory in order to bring the parameter values into comparison (Table 4).

Table 5 Comparisons between the *R* square values of four possible models for the area-perimeter relationships of the cities and towns in Yangtze River Delta, China (1985-2005)

| Year | Linear model | Exponential model | Logarithmic model | Power model |
|---|---|---|---|---|
| 1985 | 0.747 | 0.372 | 0.793 | **0.928** |
| 1996 | **0.919** | 0.441 | 0.722 | 0.914 |
| 2005 | **0.939** | 0.439 | 0.626 | 0.888 |

Second, both the boundary dimension and the form dimension approach certain constants. This seems to suggest some conversation law of fractal parameter evolution. The average revised boundary dimension approaches to a constant: $D_b$=1.182 for 1985, $D_b$=1.236 for 1996, and $D_b$=1.227 for 2005. In short, the mean of the textual dimension is close to 1.2, namely, $D_b \to 1.2$. Accordingly, the average form dimension approaches another constant: $D_f$=1.707 for 1985, $D_f$=1.693 for 1996, and $D_f$=1.696 for 2005. In short, the mean of the structural dimension is close to 1.7, namely, $D_f \to 1.7$ (Table 4). The average value of the form dimension lend further support to the suggestion that the structural dimension of urban urban changes around 1.7 (Batty and Longley, 1994; Chen, 2013). The $D_f$=1.7 is an interesting value for the form dimension of cities, and the corresponding boundary dimension is about $D_b$=1.2.

Third, the fractal models and the fractal parameters reflect the social and economic state of China. For a long time, China is socialistic country based on centrally planned economy rather than market economics. Chinese city development is always associated with its political and economical conditions. Since the introduction of the policies of reform and opening-up at the end of 1978 and with the gradual establishment of a socialist market economic system from 1992, namely, after Deng's South Tour Speeches, the top-down command economics and the bottom-up market economics combined with one another to form a mixed economics (Table 6). Chinese national economy and cities developed rapidly (Yang, 1998). If we relate the urban evolution with the political and economical background of China, we can understand the changes of fractals and fractal dimension of Chinese cities. At the beginning of reform and opening up of China, i.e, in 1985, the fractal dimension values is chaotic to some extent, and several values are abnormal: the $D_b$ value is less than 1 or even less than 0, and the $D_f$ value is great than 2. The $D_b$ values range from -0.330 (an abnormal value) to 1.525, and $D_f$ values vary from 0.398 to 2.053. However, in 2005, both the $D_b$ and $D_f$ values fluctuate between 1 and 2. The $D_b$ values range from 1.046 to 1.422, and $D_f$ values vary from 1.542 to 1.916 (Table 3). The boundary dimension of cities is proved to be a function of the compactness of urban form. The fractal evolution from chaos to order can be mirrored by the relations between the boundary dimension and the compactness ratio(Figure 3).

Table 6 Important historical events associated with urban evolution of China

| Chance/Change | Time | Mark/Event | Consequence |
| --- | --- | --- | --- |
| Chinese economic reform and open-up | 1978-12-18 1978-12-22 | Chinese eleventh CPC Central Committee Third Plenary Session | Close economic systems change to open systems |
| The socialist market economic system | 1992-1-18 1992-2-21 | Deng's South Tour Speeches | Self-organized economics appears |
| Further economic reform and open-up | 2001-12-11 | Joining World Trade Organization (WTO) | Introduced international rules into open economic systems |

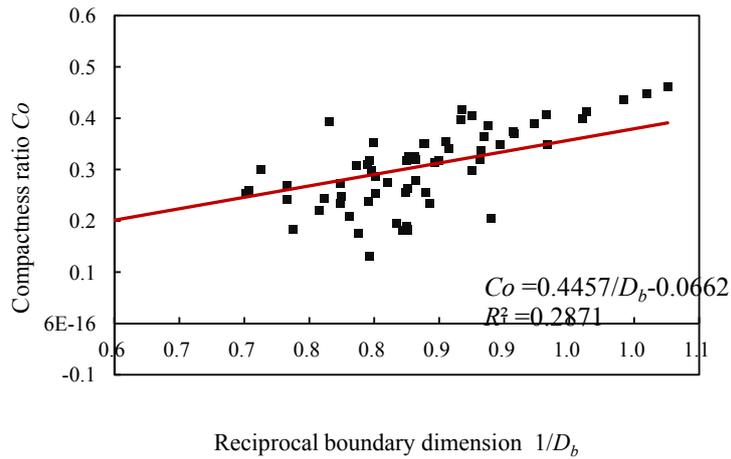

a. 1985

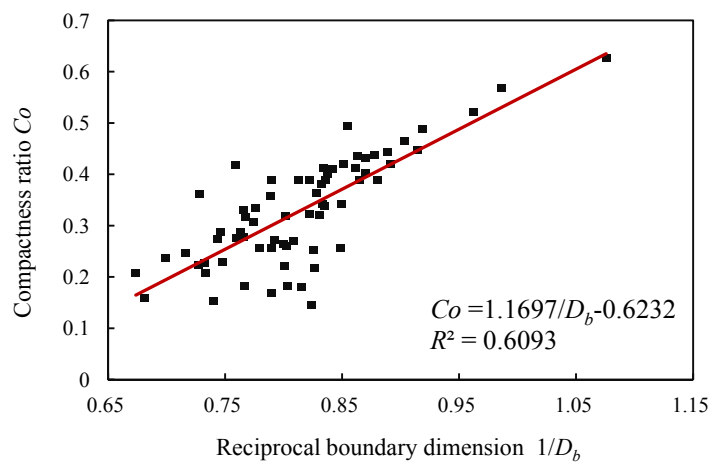

b. 1996

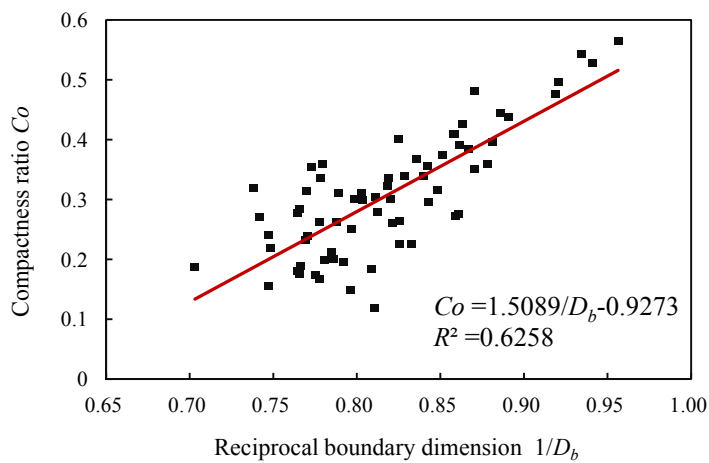

c. 2005

**Figure 3 The relationships between reciprocal of the boundary dimension and the compactness ratio of the cities and towns in Yangtze River Delta, China (1985-2005)** [**Note**: For 1985, the fractal dimension of Deqing is less than 0 and is removed as an outlier]

## 4 Questions and discussion

An urban analytical process based on the area-perimeter scaling is illustrated by using the 68

cities and towns in the study area. Through the empirical analysis, the following questions have been clarified. First, the fractal dimension of urban shape includes *form dimension* and *boundary dimension*, which are associated with one another. However, the boundary dimension of cities and patches of a city are indeed overestimated by using the traditional area-perimeter scaling formula. In Table 3, many $D_l$ values are close to 2, and this is unreasonable because the dimension of a fractal boundary is often less than 3/2. The boundary dimension used to be expressed as $D_l=bd=2b$ according to equation (1). Here $d=2$ represents the Euclidean dimension of urban region. This suggests that the urban area is regarded as a 2-dimensional measure. However, the average dimension of urban area in a system is not actually equal to 2 (Benguigui *et al*, 2006; Chen, 2013; Cheng, 1995; Imre and Bogaert, 2004). According to equation (5), the revised boundary dimension can be expressed as $D_b=\sigma D_f$, where the form dimension is near 1.7 where average is concerned (Batty and Longley, 1994; Chen, 2010a). Second, using the adjusting formula of fractal dimension, we can not only revise the boundary dimension indicating *urban texture*, but also estimate the form dimension indicative of *urban structure*. The two formulae were previously derived (Chen, 2013), but the systematic empirical analysis is made for the first time. Third, in terms of the area-perimeter scaling, the fractal dimensions of urban shape fall into two types: the *global parameters* indicative of a system of cities and the *local parameter* indicating the individual cities in the urban system. If we estimate the global parameters by using the least squares calculations, the proportionality coefficient values in the local fractal parameter formula can be obtained by the constants of the global models. In other words, if we estimate the $k$ values for an urban system by the regression analysis based on equation (1), we can compute the boundary dimension of each city in the urban system by substituting the $k$ value in equation (3). The analytical process of fractal cities can be demonstrated by a block diagram (Figure 4).

The main limitation of this study lies in the resolution of the remote sensing images. The resolution of the remote sensing images of the cities and towns are not high enough to guarantee the spatial data quality of individual cities. As a result, we cannot calculate the boundary dimension and form dimension for each city by using the least squares method. What we can do is to estimate the fractal parameters for the urban systems by means of the area-perimeter scaling based on cross-sectional data, and then estimate the local fractal parameters for each city by the approximate formula. The shortcomings of the empirical analysis are as below. First, there exist many outliers in the results. For example, the boundary dimension of Deqing city is a negative, which is absurd as a fractal dimension must be greater than 0 and less than the Euclidean dimension of the embedding space. A fractal city based on the remote sensing images is actually defined in a 2-dimension space (Batty and Longley, 1994). Thus the Euclidean dimension of the embedding space for the fractal city is $d=2$. The boundary dimension of Shangyu is greater than 2, going beyond the Euclidean dimension of its embedding space. Second, the comparability of the fractal dimension values in different year is doubtful. Generally speaking, the form dimension increases from year to year until the limit ($D_{max}=d=2$) due to space filling. However, what with the data quality and what with the urban sprawl, the form dimension values of the cities and towns randomly fluctuated in this case. Third, the numerical relation between the boundary and the compactness ratio degenerated. The boundary dimension of a city can be associated with the compactness of its urban form. The mathematical relation between the compactness ratio and the boundary dimension can be derived as below (Chen, 2011)

$$Co = \frac{K}{P}\exp(\frac{1}{D}\ln P), \qquad (10)$$

where $Co$ denotes the compactness ratio of urban form, $D$ represents the boundary dimension ($D_l$ or $D_b$), $P$ is the length of urban perimeter, and $K$ is a coefficient. However, based on the datasets above displayed, the exponential form of equation (10) is replaced by a hyperbolic such as

$$Co = u + \frac{v}{D}, \qquad (11)$$

where $u$ and $v$ represent the intercept and slope if the reciprocal of the fractal dimension, $1/D$, is treated as an independent variable. Despite these shortcomings, we can make use of the advantages and bypass the disadvantages of the remote sensing data in virtue of the formulae of the fractal dimension estimation.

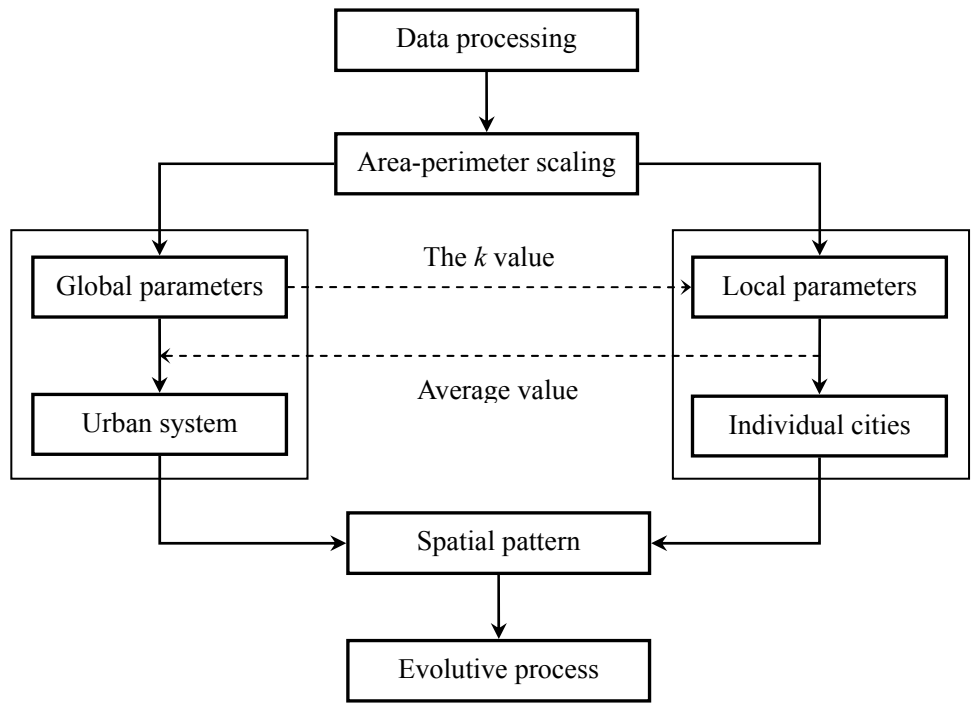

**Figure 4 A schematic diagram of the urban analytical process based on the area-perimeter scaling**

A popular wrong understanding of the boundary dimension should be pointed here. In previous literature, the initial boundary dimension $D_l$ is always associated with the stability index of urban evolution (Zhu, 2007). If the parameter $D_l$=1.5, the urban boundary is treated as resulting from Brownian motion and thus the urban development is considered to be unstable. This viewpoint is wrong. The reasons are as follows: First, it is the self-affine record dimension rather than the self-similar trail dimension that can be directly related with Brownian motion (Feder, 1988). However, the boundary dimension of urban shape is a self-similar trail dimension instead of a self-affine record dimension (Chen, 2010b). Second, as indicated above, the boundary dimension is always numerically overestimated by the old formula. In fact, the boundary dimension is seldom greater than or equal to 1.5. An urban boundary can be treated as a fractal line, and the dimension of a fractal line is usually less than 1.5. The boundary dimension should be revised using the adjusting formula, equation (6). The revised boundary dimension values often come between 1 and 1.5. Finally, a random process does not indicate an unstable process. If the self-affine record dimension is close to 1.5, it suggests randomicity rather than instability of urban growth.

## 5 Conclusions

This work tries to develop an approach to make the best of the advantages and bypass the disadvantages of spatial data by means of the new fractal dimension formulae. The innovation of this article rests with two aspects: First, the global fractal parameters and the local fractal parameters are integrated to make urban spatial analyses, and the numerical link between the global and local parameters is revealed by statistical average. Second, the area-perimeter scaling is employed to estimate the proportionality coefficient for the approximate formula of boundary dimension. Two findings are made in this study. One is numerical link between the global and local parameters, and the other is the degeneration phenomenon of area-perimeter scaling relation.

The main conclusions can be reached as follows: First, the relationships between urban area and perimeter follows the allometric scaling laws, the scaling exponent is the ratio of urban boundary dimension and urban form dimension. However, the boundary dimension value used to be overestimated because of the form dimension ($D_f$<2) is mistaken for Euclidean dimension of urban area ($d$=2). By using the revision formulae of fractal dimension, we can correct the results of fractal dimension estimation. What is more, the form dimension can be estimated through the boundary dimension. Thus, the form dimension indicative urban structure and the boundary dimension indicating urban texture can be combined with each other to characterize urban evolution. Second, a fractal system of cities should be characterized by both the global and local scaling parameters. The fractal parameters of an urban system, including form dimension and boundary dimension, fall into two groups: global parameters and local parameters. The global parameters reflect the spatial properties of a system of cities, while the local parameters reflect the geographical feature of individual cities. The average values of the local parameters are approximately equal to the corresponding values of the global parameters. This suggests that the global parameters can be decomposed into local parameters. By means of the average values, we can associate the global parameters with local parameters, and further connect the macro level of an urban system with the micro level of elements in the urban system. Third, the fractal measure relations between urban area and urban envelope may degenerate because of disorder competition of cities. Fractals follow scaling laws, which can be expressed as a set of power functions. However, a power-law relation between urban area and perimeter sometimes degenerates into a semilogarithmic relation or even a linear relation. Fractal structure based on power laws is a kind of optimized structure of natural and human systems. A fractal object can fill its space in the best way. Fractal relation degeneration suggests some latent problems of urban evolution. If a city or urban system does not follow the allometric scaling law, the fractal structure may be broken and the structure of the city or system of cities should be improved by scientific city planning.

## Acknowledgement

This research was supported financially by the National Natural Science Foundation of China (Grant no. 41171129). The support is gratefully acknowledged.

## References

Addison PS (1997). *Fractals and Chaos: An Illustrated Course*. Bristol and Philadelphia: Institute of Physics Publishing

Batty M (2008). The size, scale, and shape of cities. *Science*, 319: 769-771

Batty M, Carvalho R, Hudson-Smith A, Milton R, Smith D, Steadman P (2008). Scaling and allometry in the building geometries of Greater London. *The European Physical Journal B - Condensed Matter and Complex Systems*, 63(3): 303-314

Batty M, Longley PA (1988). The morphology of urban land use. *Environment and Planning B: Planning and Design*, 15(4): 461-488

Batty M, Longley PA (1989). On the fractal measurement of geographical boundaries. *Geographical Analysis*, 21(1): 47-67

Batty M, Longley PA (1994). *Fractal Cities: A Geometry of Form and Function*. London: Academic Press

Batty M, Xie Y (1999). Self-organized criticality and urban development. Discrete Dynamics in Nature and Society, 3(2-3): 109-124

Benguigui L, Blumenfeld-Lieberthal E, Czamanski D (2006). The dynamics of the Tel Aviv morphology. *Environment and Planning B: Planning and Design*, 33: 269-284

Benguigui L, Czamanski D, Marinov M (2001). The dynamics of urban morphology: the case of Petah Tikvah. *Environment and Planning B: Planning and Design*, 28: 447-460

Benguigui L, Czamanski D, Marinov M, Portugali Y (2000). When and where is a city fractal? *Environment and Planning: Planning and Design*, 27(4): 507–519

Bettencourt LMA (2013). The origins of scaling in cities. *Science*, 340: 1438-1441

Bettencourt LMA, Lobo J, Helbing D, Kühnert C, West GB (2007). Growth, innovation, scaling, and the pace of life in cities. *Proceeding of the National Academy of Sciences of the United States of America*, 104(17): 7301-7306

Chang XL (1996). The study of relationship between the process of desertification and the landscape pattern in Bashing region, Hebei Province. *Journal of Desert Research*, 16: 222-227 (in Chinese)

Chang XL, Wu JG (1998). Spatial analysis of pattern of sandy landscapes in Kerqin, Inner Mongolia. *Acta Ecologica Sinica*, 18(3): 225-232 (in Chinese)

Chen T (1995). *Studies on Fractal Systems of Cities and Towns in the Central Plains of China (Master's Degree Thesis)*. Changchun: Department of Geography, Northeast Normal University [in Chinese]

Chen YG (2010a). Characterizing growth and form of fractal cities with allometric scaling exponents. *Discrete Dynamics in Nature and Society*, vol. 2010, Article ID 194715, 22 pages

Chen YG (2010b). Exploring the fractal parameters of urban growth and form with wave-spectrum analysis. *Discrete Dynamics in Nature and Society*, vol. 2010, Article ID 974917, 20 pages

Chen YG (2011). Derivation of the functional relations between fractal dimension and shape indices of urban form. *Computers, Environment and Urban Systems*, 35(6): 442–451

Chen YG (2013). A set of formulae on fractal dimension relations and its application to urban form. *Chaos, Solitons & Fractals*, 54(1): 150-158

Chen YG, Jiang SG (2009). An analytical process of the spatio-temporal evolution of urban systems based on allometric and fractal ideas. *Chaos, Solitons & Fractals*, 39(1): 49-64

Chen YG, Wang JJ (2013). Multifractal characterization of urban form and growth: the case of Beijing. *Environment and Planning B: Planning and Design*, 40(5):884-904

Cheng Q (1995). The perimeter-area fractal model and its application in geology. *Mathematical Geology*, 27 (1): 69-82

De Keersmaecker M-L, Frankhauser P, Thomas I (2003). Using fractal dimensions for characterizing intra-urban diversity: the example of Brussels. *Geographical Analysis*, 35(4): 310-328

Feder J (1988). *Fractals*. New York: Plenum Press

Feng J, Chen YG (2010). Spatiotemporal evolution of urban form and land use structure in Hangzhou, China: evidence from fractals. *Environment and Planning B: Planning and Design*, 37(5): 838–856

Frankhauser P (1994). *La Fractalité des Structures Urbaines (The Fractal Aspects of Urban Structures)*. Paris: Economica (In French)

Frankhauser P (1998). The fractal approach: A new tool for the spatial Analysis of urban agglomerations. *Population: An English Selection*, 10(1): 205-240 [New Methodological Approaches in the Social Sciences]

Gordon K (2005). The mysteries of mass. *Scientific American*, 293(1):40-46/48

Henry J (2002). *The Scientific Revolution and the Origins of Modern Science (2nd Edition)*. New York: Palgrave

Imre AR (2006). Artificial fractal dimension obtained by using perimeter-area relationship on digitalized images. *Applied Mathematics and Computation*, 173 (1): 443-449

Imre AR, Bogaert J (2004). The fractal dimension as a measure of the quality of habitat. *Acta Biotheoretica*, 52(1): 41-56

Kaye BH (1989). *A Random Walk Through Fractal Dimensions*. New York: VCH Publishers

Lobo J, Bettencourt LMA, Strumsky D, West GB (2013). Urban scaling and the production function for cities. *PLoS ONE*, 8(3): e58407

Longley PA, Batty M (1989). Fractal measurement and line generalization. *Computer & Geosciences*, 15(2): 167-183

Longley PA, Batty M, Shepherd J (1991). The size, shape and dimension of urban settlements. *Transactions of the Institute of British Geographers (New Series)*, 16(1): 75-94

Luo HY, Chen YG (2014). An allometric algorithm for fractal-based Cobb-Douglas function of geographical systems. *Discrete Dynamics in Nature and Society*, vol. 2014, Article ID 910457, 10 pages

Mandelbrot BB (1983). *The Fractal Geometry of Nature*. New York: W. H. Freeman and Company

Shen G (2002). Fractal dimension and fractal growth of urbanized areas. *International Journal of Geographical Information Science*, 16(5): 419-437

Song Y, Wang SJ, Ye Q, Wang XW (2012). Urban spatial morphology characteristic and its spatial differentiation of mining city in China. *Areal Research and Development*, 31(1):45-39 (In Chinese)

Takayasu H (1990). *Fractals in the Physical Sciences*. Manchester: Manchester University Press

Thomas I, Frankhauser P, Biernacki C (2008). The morphology of built-up landscapes in Wallonia (Belgium): A classification using fractal indices. *Landscape and Urban Planning*, 84(2): 99-115

Thomas I, Frankhauser P, De Keersmaecker M-L (2007). Fractal dimension versus density of built-up surfaces in the periphery of Brussels. *Papers in Regional Science*, 86(2): 287-308

Thomas I, Frankhauser P, Frenay B, Verleysen M (2010). Clustering patterns of urban built-up areas with curves of fractal scaling behavior. *Environment and Planning B: Planning and Design*, 37(5): 942-954

Wang XS, Liu JY, Zhuang DF, Wang Liming. (2005) Spatial-temporal changes of urban spatial morphology in China. *Acta Geographica Sinica*, 60(3): 392-400 (In Chinese)

White R (1998). Cities and cellular automata. *Discrete Dynamics in Nature and Society*, 2(2): 111-125

White R, Engelen G (1994). Urban systems dynamics and cellular automata: fractal structures between order and chaos. *Chaos, Solitons & Fractals*, 4(4): 563-583

Yang DL (1998). *Calamity and Reform in China: State, Rural Society, and Institutional Change since the Great Leap Famine*. Redwood: Stanford University Press

Zhu XH (2007). *Fractal and Fractal Dimensions of Spatial Geo-Information*. Beijing: Surveying and Mapping Press (In Chinese)